\documentclass{PoS}

\usepackage{wrapfig}

\title{Hessian PDF reweighting meets the Bayesian methods}

\ShortTitle{Hessian PDF reweighting meets the Bayesian methods}

\author{\speaker{Hannu Paukkunen} \\
        Department of Physics, University of Jyv\"askyl\"a, P.O. Box 35, \\ FI-40014 University of Jyv\"askyl\"a, Finland  \\ 
        Helsinki Institute of Physics, University of Helsinki, P.O. Box 64, FI-00014, Finland \\
        E-mail: \email{hannu.paukkunen@jyu.fi}}

\author{{Pia Zurita} \\
        Departamento de F\'\i sica de Part\'\i culas and IGFAE, Universidade de Santiago de
Compostela, E-15782 Galicia, Spain\\
        E-mail: \email{pia.zurita@usc.es}}
        
\abstract{We discuss the Hessian PDF reweighting --- a technique intended to estimate the effects that new measurements
have on a set of PDFs. The method stems straightforwardly from considering new data in a usual $\chi^2$-fit and it naturally
incorporates also non-zero values for the tolerance, $\Delta\chi^2>1$. In comparison to the contemporary Bayesian reweighting techniques,
there is no need to generate large ensembles of PDF Monte-Carlo replicas, and the observables need to be evaluated only with 
the central and the error sets of the original PDFs. In spite of the apparently rather different methodologies, we find that
the Hessian and the Bayesian techniques are actually equivalent if the $\Delta\chi^2$ criterion is properly 
included to the Bayesian likelihood function that is a simple exponential.}

\FullConference{XXII. International Workshop on Deep-Inelastic Scattering and Related Subjects,\\
		28 April - 2 May 2014\\
		Warsaw, Poland}

\begin{document}

\section{Introduction}

The flood of hard-process data from the LHC proton+proton collisions that can
test and offer further constraints for the parton distribution functions (PDFs) 
is nowadays so massive that the need to efficiently quantify the implications of different
measurements has called for novel analysis techniques. To this end, an option that 
has gained some popularity is to make use of HERAFitter \cite{SAPRONOV:2014qla} to
check the constraining power of the new data. However, in most cases this has meant
comparing the PDFs obtained by using only the HERA deep-inelastic scattering data
with the ones including additionally a specific set of LHC data (see e.g. \cite{Chatrchyan:2013mza,CMS:2013yua}). 
Clearly, there is no guarantee that this would reflect the impact of the new data in the global
context. In this case the PDF reweighting methods
\cite{Ball:2010gb,Ball:2011gg,Watt:2012tq,Paukkunen:2013grz,Paukkunen:2014zia}, discussed in this talk,
should be more adequate.

\section{The Hessian reweighting}
\label{sec:Hess}

Let us suppose we have a set of Hessian PDFs with a global tolerance $\Delta\chi^2$.
The PDFs have been parametrized by some fixed functional form and the central set $S_0$
corresponds to those parameter values $a^0_i$ that minimize a global $\chi^2$-function.
The Hessian procedure \cite{Pumplin:2001ct} to quantify the PDF errors is based
on expanding this $\chi^2$-function around the minimum $\chi^2_0$ with respect to 
the fit parameters $a_i$ and diagonalizing the Hessian matrix $H_{ij}$:
\begin{equation}
 \chi^2\{a\} \approx \chi^2_0 + \sum_{ij} (a_i-a_i^0) H_{ij} (a_j-a_j^0)
 = \chi^2_0  + \sum_i  z_i^2. \label{eq:chi2orig}
\end{equation}
The coordinates of the central set $S_0$ and error sets $S^\pm_k$ in this $z$-space 
(``space of eigenvectors'') are
\begin{eqnarray}
z({S_0}) & = & \left(0,0,...,0 \right), \nonumber \\
z({S^\pm_1}) & = & \pm \sqrt{\Delta \chi^2} \left(1,0,...,0 \right), \label{eq:errset} \\
z({S^\pm_2}) & = &  \pm \sqrt{\Delta \chi^2} \left(0,1,...,0 \right). \nonumber \\
         & \vdots & \nonumber
\end{eqnarray}
The idea elaborated in Refs.~\cite{Paukkunen:2013grz,Paukkunen:2014zia} is
to add the contribution of a new set of data $\{y\}$ with covariance matrix $C_{ij}$ to the Eq.~(\ref{eq:chi2orig}) above
\begin{equation}
 \chi^2_{\rm new} \equiv \chi^2_0  +   \sum_k z_k^2  + 
 \sum_{i,j} \left(y_i[f]-y_i\right) C_{ij}^{-1} \left(y_j[f]-y_j\right), \label{eq:newchi2}
\end{equation}
and estimate the PDF-dependent theory values $y_i[f]$ by a linear approximation as
\begin{equation}
 y_i \left[f \right] \approx y_i \left[{S_0} \right] + \sum_{k} \frac{\partial y_i \left[{S} \right]}{\partial z_k}{\Big|_{S=S_0}} z_k
                   \approx y_i \left[S_0 \right] + \sum_{k} D_{ik} w_k, \label{eq:XS}
\end{equation}
where 
\begin{equation}
D_{ik} \equiv \frac{y_i\left[S_k^+ \right] - y_i\left[S_k^- \right]}{2} \quad {\rm and} \quad
w_k    \equiv \frac{z_k}{\sqrt{\Delta \chi^2}}. 
\end{equation}
The function $\chi^2_{\rm new}$ is thus a second-order polynomial in variables $w_i$ and
its minimum occurs at
$
{\vec {\bf w}^{\rm min}} = -{\bf B}^{-1} \vec {\bf a}. \label{eq:wmineq} 
$
with
\begin{equation}
 B_{kn} = \sum_{i,j} {D_{ik} C_{ij}^{-1} D_{jn}} + \Delta \chi^2 \delta_{kn} \, , \quad
 a_k    = \sum_{i,j} {D_{ik} C_{ij}^{-1} \left( y_j\left[S_0\right] - y_j \right)}.
\end{equation}
The corresponding new PDFs $f^{\rm new}$ (omitting here all arguments and flavor indices) are easily obtained, by the same approximation as in Eq.~(\ref{eq:XS}),
\begin{equation}
 f^{\rm new} \approx f_{S_0} + \sum_{k} \left( \frac{f_{S^+_k}-f_{S^-_k}}{2} \right) w^{\rm min}_k, \label{eq:newPDF}
\end{equation}
and by rewriting the function $\chi^2_{\rm new}$ as
$
 \chi^2_{\rm new} = \chi^2_{0,{\rm new}}  +  \sum_{ij} \delta w_i B_{ij} \delta w_j, 
$
one can also define the new PDF error sets by the same procedure as above. The increase of the original $\chi^2$ ---
the ``reweighting penalty'' --- can be approximated by
\begin{equation}
P \approx \Delta \chi^2 \sum_{k=1} (w^{\rm min}_{k})^2.
\end{equation}

\section{The Bayesian procedures}

The Bayesian PDF reweighting methods data back to the original works of Giele and Keller \cite{Giele:1998gw}
and were later on revived by the NNPDF collaboration \cite{Ball:2010gb,Ball:2011gg}. To apply these techniques
in the case of Hessian PDFs one constructs an ensemble of PDF replicas by \cite{Watt:2012tq}
\begin{equation}
 f_k \equiv f_{S_0} + \sum_i \left( \frac{f_{S^+_i}-f_{S^-_i}}{2} \right) R_{ik} \label{eq:replicas},
\end{equation}
where $R_{ik}$ are Gaussian random numbers. The PDF dependent observables are then obtained
as expectation values
\begin{equation}
\langle \mathcal{O} \rangle  = \frac{1}{N_{\rm rep}} \sum_{k=1}^{N_{\rm rep}} \mathcal{O} \left[ f_k \right], \\
\end{equation}
which coincide with $\left[ f_{S_0} \right]$ if the non-linearities are small and the number of replicas $N_{\rm rep}$
is sufficiently large. The Bayesian reweighting amounts to turning these averages to weighted ones
\begin{equation}
\langle \mathcal{O} \rangle_{\rm new}  =  \frac{1}{N_{\rm rep}} \sum_{k=1}^{N_{\rm rep}} \omega_k \, \mathcal{O} \left[ f_k \right], \label{eq:Bnew} 
\end{equation}
where the weights $\omega_k$ are determined solely from the new data. Two different functional forms have
appeared in the literature: The one proposed originally by Giele and Keller is a simple exponential
\begin{equation}
 \omega_k^{\rm GK} = \frac{\exp\left[-\chi^2_k/2 \right]}{(1/N_{\rm rep}) \sum_{k=1}^{N_{\rm rep}} 
\exp\left[-\chi^2_k/2 \right]}, \label{eq:wGK}
\end{equation}
and the one that has been explicitly shown \cite{Ball:2010gb,Ball:2011gg} to work with the NNPDF fit
framework resembles a chi-squared distribution
\begin{equation}
 \omega_k^{{\rm chi-squared}} = \frac{\left( \chi^2_k \right)^{(N_{\rm data}-1)/2} \exp\left[-\chi^2_k/2 \right]}{(1/N_{\rm rep}) \sum_{k=1}^{N_{\rm rep}} \left( \chi^2_k \right)^{(N_{\rm data}-1)/2} 
\exp\left[-\chi^2_k/2 \right]}. \label{eq:wNNPDF}
\end{equation}
In both,
\begin{equation}
 \chi^2_k = \sum_{i,j} \left(y_i[f_k]-y_i\right) C_{ij}^{-1} \left(y_j[f_k]-y_j\right). \label{eq:chi2onlynew}
\end{equation}
The reweighting penalty can be computed by
\begin{eqnarray}
P \approx \Delta \chi^2 \, \sum_i \left( \frac{1}{N_{\rm rep}} \sum_k^{N_{\rm rep}} \omega_k R_{ik} \right)^2. \label{eq:penaltyB}
\end{eqnarray}

\section{Simplified example}

\begin{figure}[th!]
\centering
\includegraphics[width=0.85\textwidth]{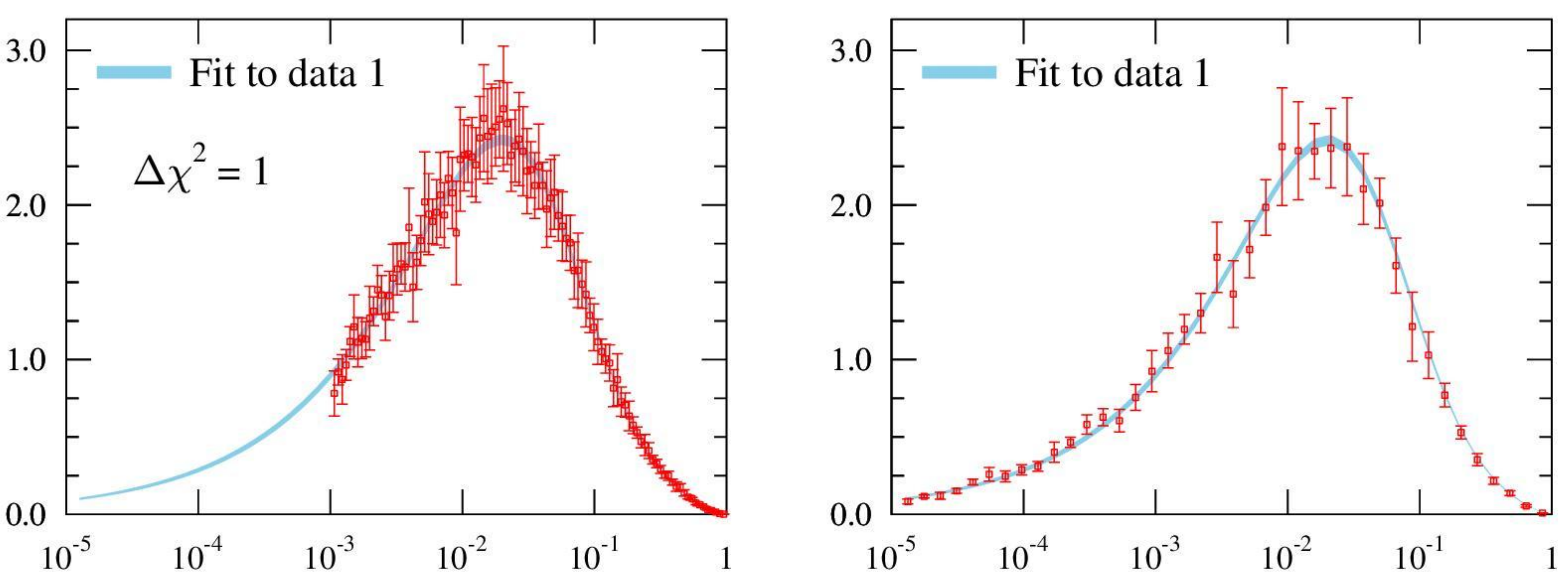}
\caption{{\bf Left-hand panel:} Pseudodata (data set 1) used to construct the baseline fit. 
{\bf Right-hand panel:} Pseudodata (data set 2) used in reweighting. } 
\label{fig:ex1}
\end{figure}

To compare the different reweighting methods we invoke a simple example by constructing
two sets of pseudodata for a function 
$
g(x) = a_0 x^{a_1} (1-x)^{a_2}  e^{x a_3}  (1 + x e^{a_4})^{a_5},
$
shown in Figure~\ref{fig:ex1}. We use the first one to construct a set of ``Hessian PDFs''
as outlined in Section~\ref{sec:Hess} (using the same functional form for $g(x)$) with a 
chosen tolerance $\Delta \chi^2$. Then we take
the second data set, work out the predictions of the reweighting methods and compare those to
a direct fit including both of these data sets.
\begin{figure}[th!]
\centering
\includegraphics[width=0.32\textwidth]{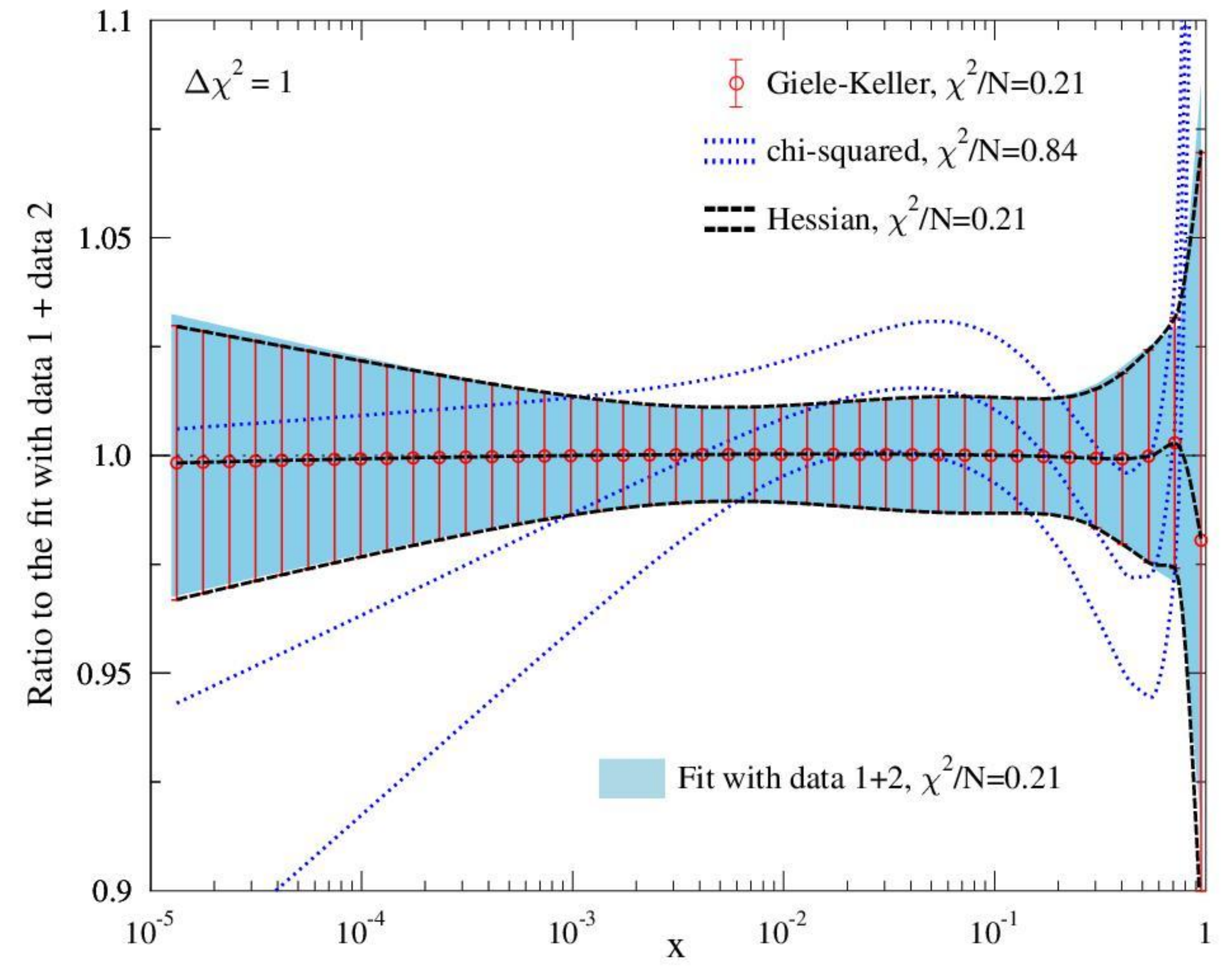}
\includegraphics[width=0.32\textwidth]{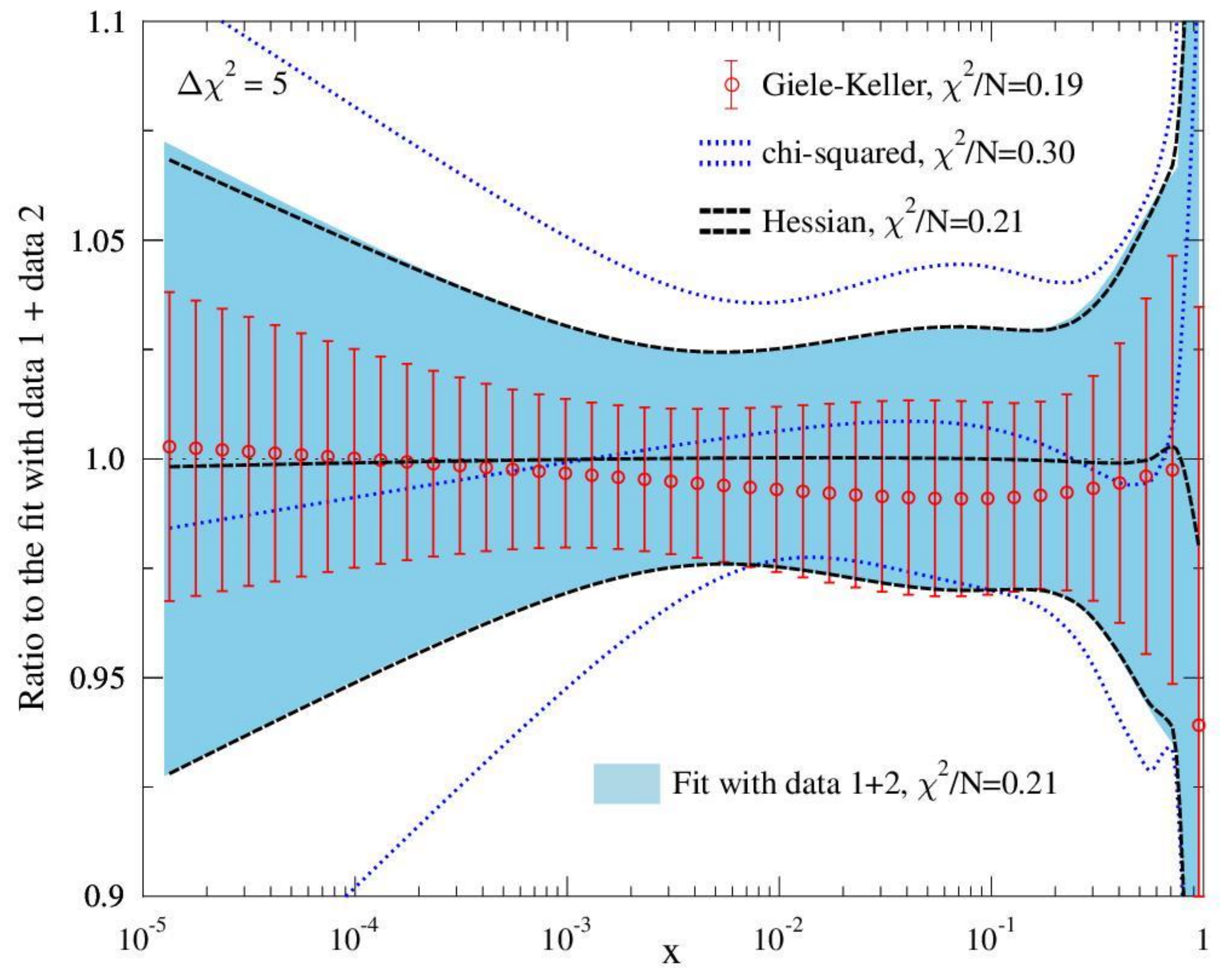}
\includegraphics[width=0.32\textwidth]{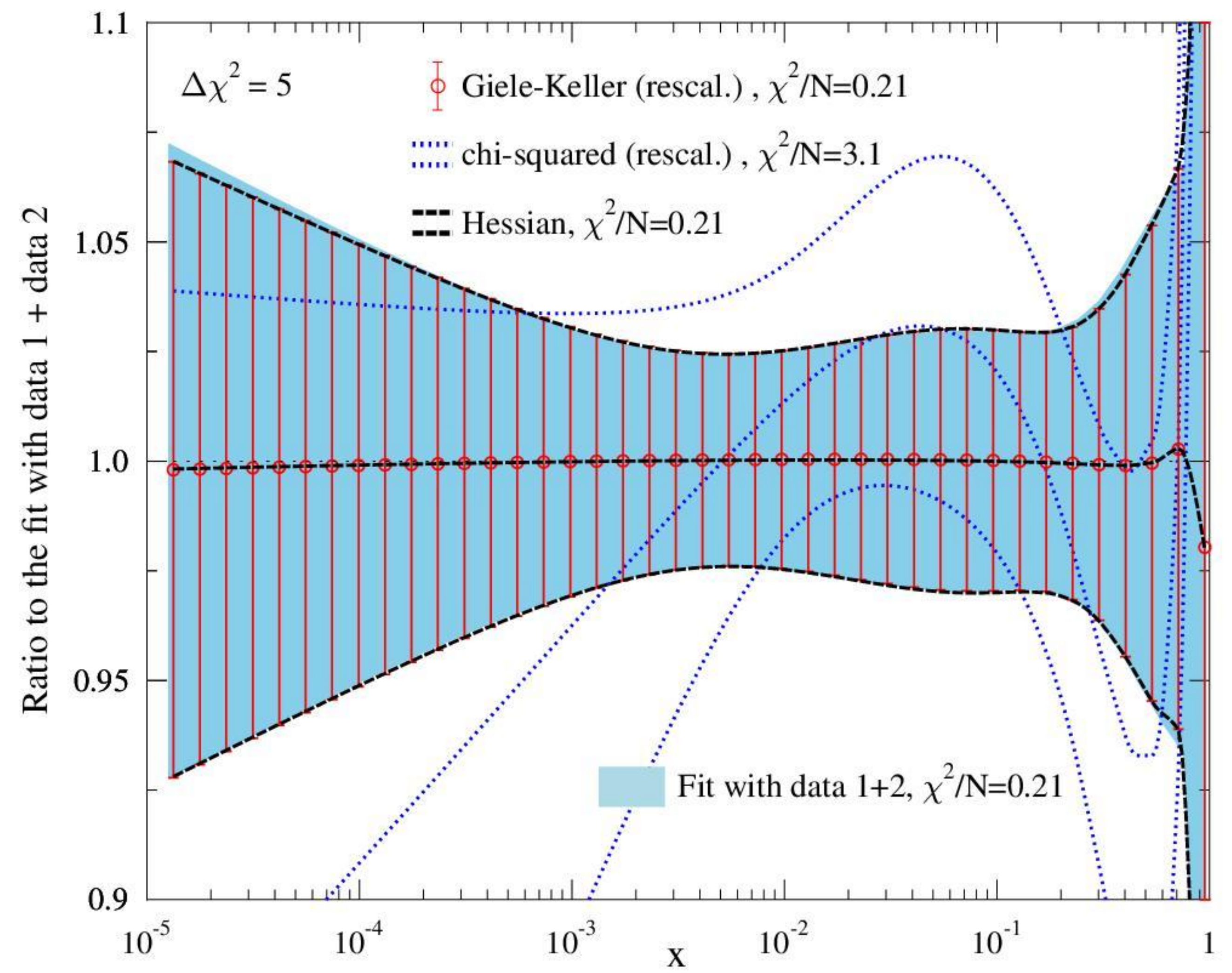}
\caption{{\bf Left-hand panel:} Results of reweighting normalized to to the direct re-fit
in the case $\Delta \chi^2=1$. {\bf Middle panel:} As the left-hand panel but for $\Delta \chi^2=5$.
{\bf Right-hand panel:} As the middle panel but rescaling by $\Delta \chi^2=5$ when computing
the Bayesian weights.} 
\label{fig:ex2}
\end{figure}
The results of this exercise are shown in Figure~\ref{fig:ex2}.
We observe that in the case $\Delta \chi^2=1$ the Hessian and Giele-Keller reweighting are in
perfect agreement with the direct fit (left-hand panel). If we increase the tolerance to $\Delta \chi^2=5$, the Hessian
procedure still accords with the direct fit but the Giele-Keller method appears to fail (middle panel). However, the
agreement can be easily restored by rescaling the values of $\chi^2_k$ in Eq.~(\ref{eq:chi2onlynew}) as 
$\chi^2_k \rightarrow \chi^2_k/\Delta \chi^2$ (left-hand panel). In all cases the Bayesian weights which
have been shown to work for the NNPDF-style fits (the chi-squared weights) yield clearly different results.

\section{CTEQ6.6 and inclusive jets at the LHC}

\begin{figure}[th!]
\centering
\includegraphics[width=0.47\textwidth]{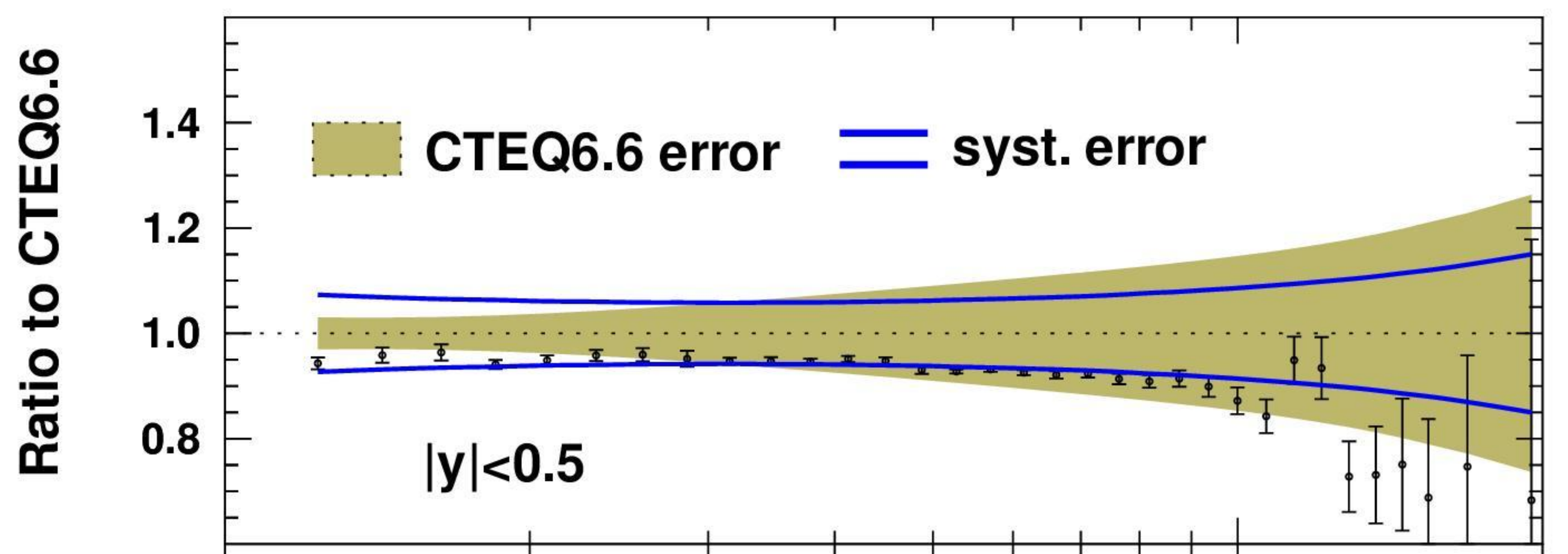}
\includegraphics[width=0.47\textwidth]{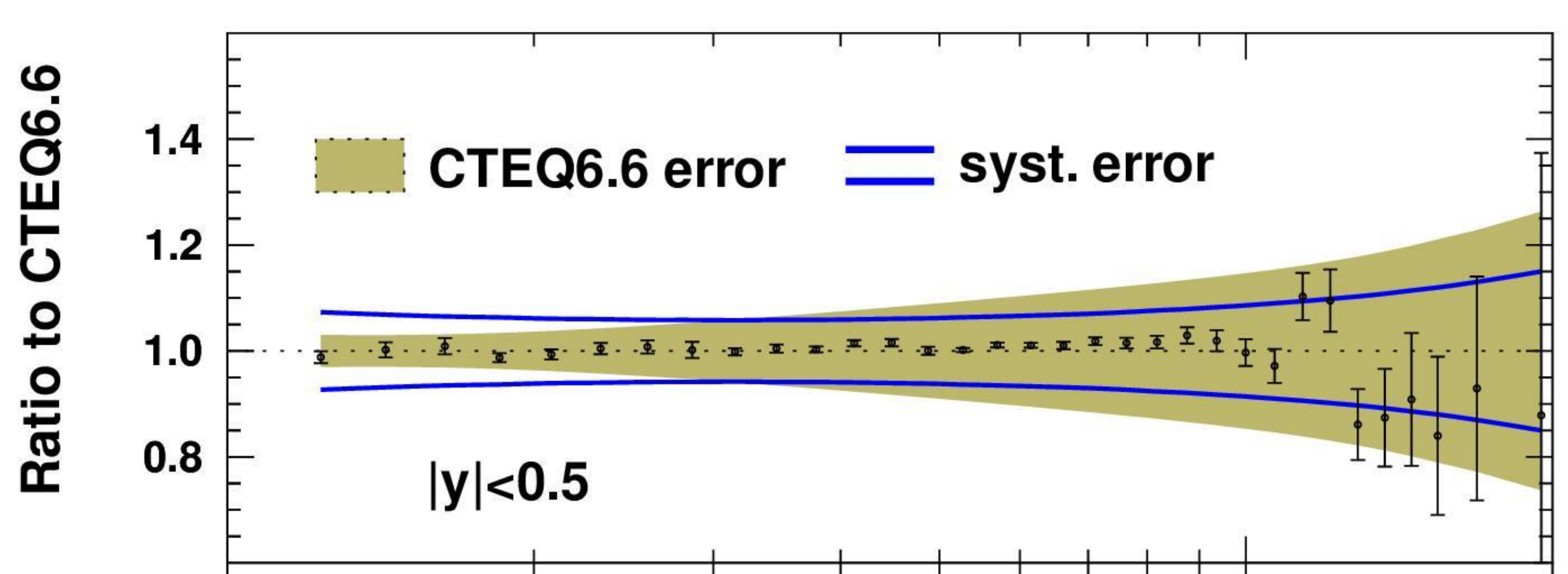}
\includegraphics[width=0.48\textwidth]{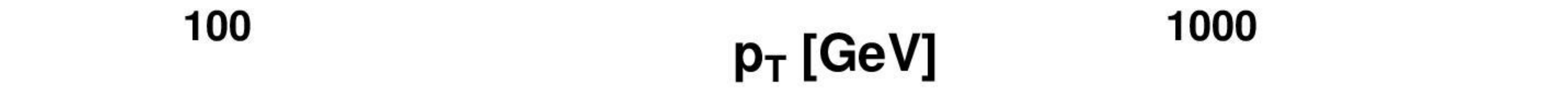}
\hspace{-0.2cm}
\includegraphics[width=0.48\textwidth]{pTaxis.pdf}
\hspace{-0.2cm}
\caption{{\bf Left-hand panel:} The CMS jet data (only the midrapidity bin) normalized by the
predictions of CTEQ6.6. {\bf Right-hand panel:} As the left-hand panel but after
applying the systematic shifts.} 
\label{fig:jet1}
\end{figure}

Having now understood how to correctly reweight Hessian PDFs, we illustrate what would be the effect of
LHC inclusive jet data on the CTEQ6.6 PDFs \cite{Nadolsky:2008zw} (for which which $\Delta \chi^2=100$). Specifically, we focus on the $7 \, {\rm TeV}$
jet measurements by the CMS collaboration \cite{Chatrchyan:2012bja} and use the FASTNLO
interface \cite{Kluge:2006xs,Britzger:2012bs,Wobisch:2011ij} for the computations. Before the reweighting
CTEQ6.6 tends to somewhat overpredict the experimental cross sections as shown in Figure~\ref{fig:jet1}
(left-hand panel) which, however, largely disappears after applying the correlated systematic shifts (right-hand panel).
Initially, $\chi^2/N \approx 2.1$ (for $N=133$ data points).

\begin{wrapfigure}{r}{0.55\textwidth}
\vspace{-0.2cm}
\centerline{\includegraphics[width=0.55\textwidth]{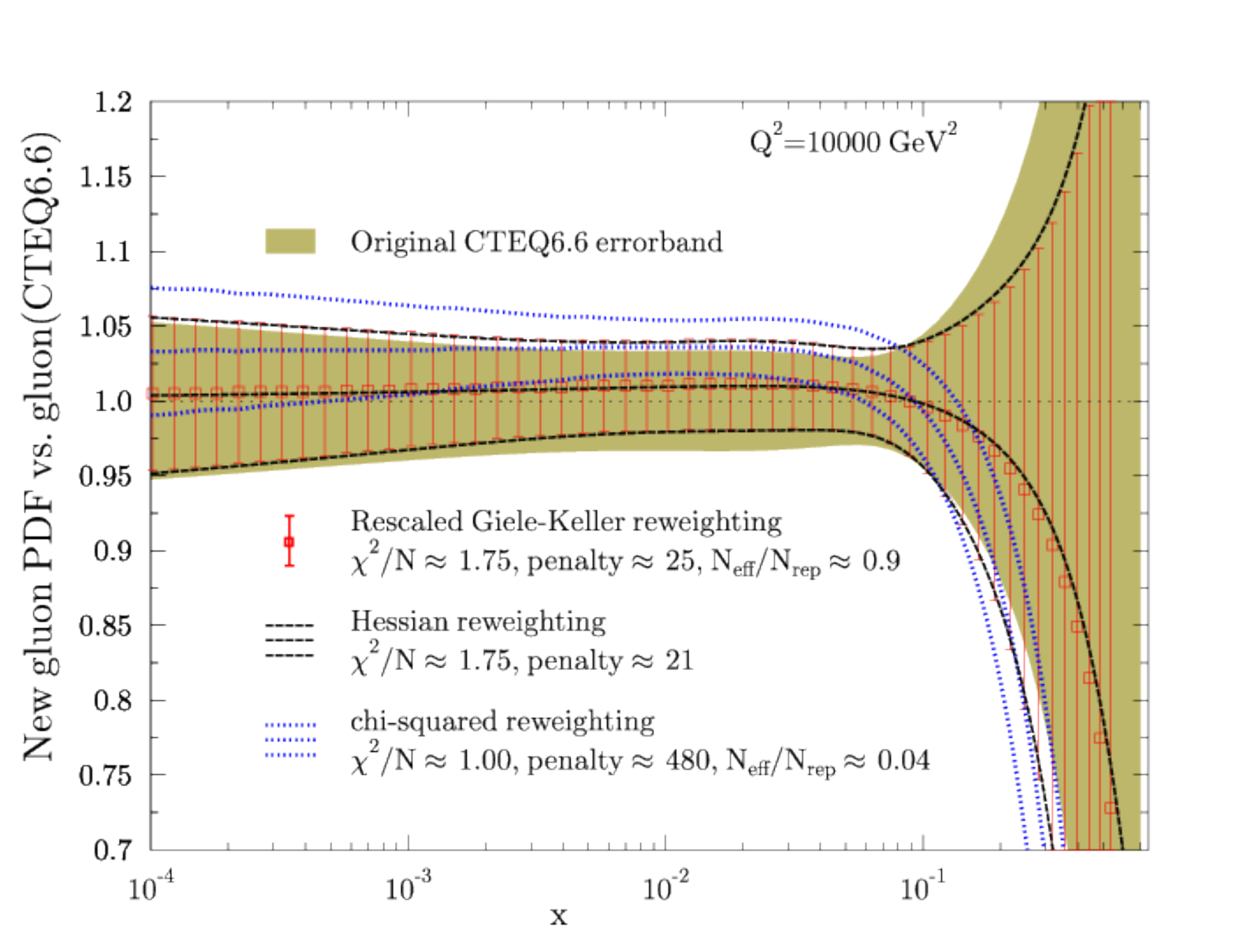}}
\caption[]{The new gluon PDFs normalized to CTEQ6.6.}
\label{fig:jet3}
\vspace{-0.2cm}
\end{wrapfigure}

\begin{figure}[th!]
\centering
\includegraphics[width=0.47\textwidth]{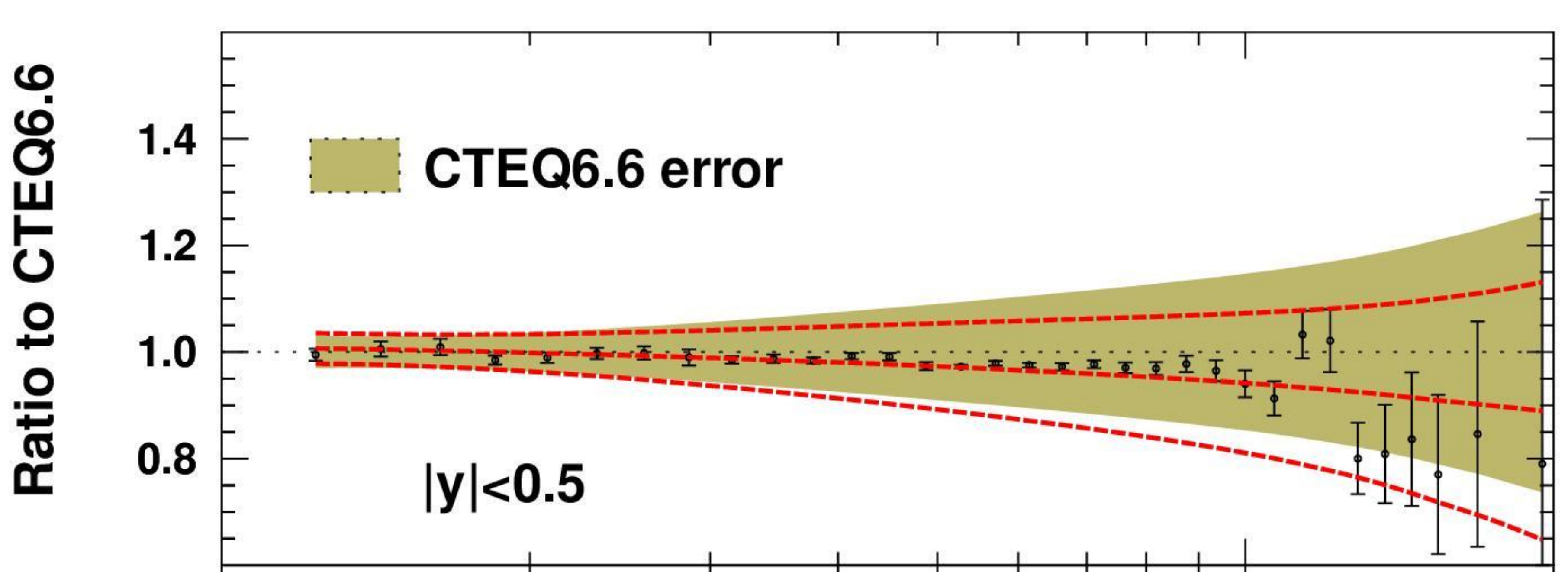}
\includegraphics[width=0.47\textwidth]{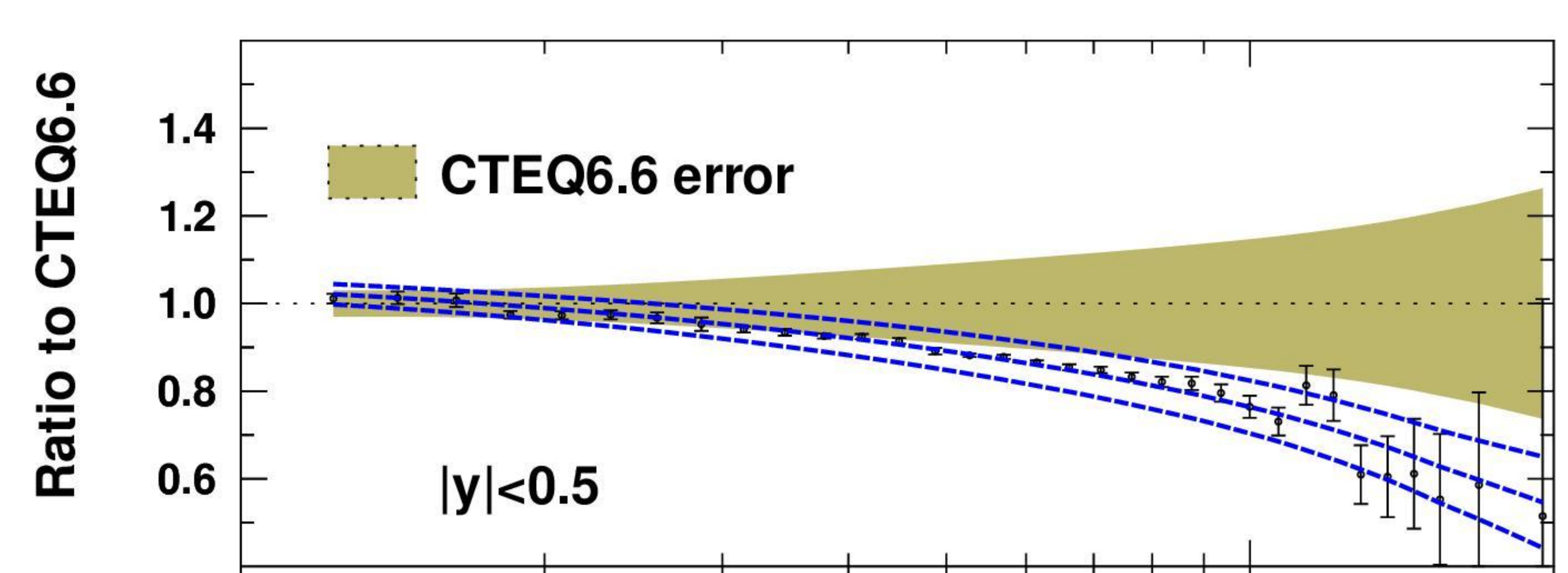}
\includegraphics[width=0.48\textwidth]{pTaxis.pdf}
\hspace{-0.2cm}
\includegraphics[width=0.48\textwidth]{pTaxis.pdf}
\hspace{-0.2cm}
\caption{{\bf Left-hand panel:} The jet cross sections after the Hessian reweighting
(red lines) normalized by the central CTEQ6.6 predictions. The systematic shifts 
corresponding to the reweighted PDFs have been applied to the data. {\bf Right-hand panel:}
As the left-hand panel but using the Bayesian reweighting with chi-squared weights.} 
\label{fig:jet2}
\end{figure}

The gluon distributions after applying the reweighting procedures are presented in Figure~\ref{fig:jet3}
revealing a decrease in the large-$x$ gluon PDF. As expected, the Hessian and (rescaled) Giele-Keller 
reweighting agree and only a modest penalty of $\sim~20$ units is induced. The new global $\chi^2$
has changed by $21-(2.1-1.75)\times 133 \approx -30$ units. The result of Bayesian reweighting with chi-squared weights
is shown for comparison and a similar but much too pronounced effect is observed. In fact, instead of decreasing,
the new global $\chi^2$ has increased by $480-(2.1-1.0)\times 133 \approx 330$ units. This is also reflected in 
the new cross-section predictions shown in Figure~\ref{fig:jet2}: While the Hessian reweighting 
(left-hand panel) predicts only a modest decrease in the cross sections (moderating the overshooting 
observed in Figure~\ref{fig:jet2}), the Bayesian reweighting with chi-squared weights (right-hand panel)
would lead us to believe in much larger effect.

\section{Summary}

We have discussed how to estimate the effects that a new set of data would have 
on a global Hessian PDF fit with fixed tolerance $\Delta \chi^2$. By considering a
simple example, we find that there are two alternative techniques that give essentially
the same answer and are equivalent to a direct refit: the Hessian reweighting and a Bayesian 
technique with rescaled Giele-Keller weights. As a practical example, we employed
these methods in the case of inclusive jet production at the LHC.

\section*{Acknowledgments}
H.P. wants to acknowledge the financial support from the Academy of Finland, Project No. 133005. 
P.Z. is supported by European Research Council grant HotLHC ERC-2011-StG-279579; by Ministerio 
de Ciencia e Innovaci\'on of Spain under project FPA2011-22776, and the Consolider-Ingenio 2010
Programme CPAN (CSD2007-00042); by Xunta de Galicia (GRC2013-024); and by FEDER.

\end{document}